\documentclass[conference]{IEEEtran}
\IEEEoverridecommandlockouts
\usepackage{cite}
\usepackage{amsmath,amssymb,amsfonts}
\usepackage{algorithmic}
\usepackage{graphicx}
\usepackage{textcomp}
\usepackage[font=footnotesize]{subcaption}
\DeclareCaptionFont{ieeesmall}{\footnotesize}
\usepackage{hyperref}
\usepackage{xcolor}
\def\BibTeX{{\rm B\kern-.05em{\sc i\kern-.025em b}\kern-.08em
    T\kern-.1667em\lower.7ex\hbox{E}\kern-.125emX}}

\usepackage[nolist]{acronym}

\usepackage{csquotes}
\usepackage{mathptmx}
\usepackage{color}
\usepackage{mwe}

\newcommand{\etal}[0]{\textit{et~al.}\xspace}
\newcommand{\eg}{\textit{e.g.}\xspace}
\newcommand{\ie}{\textit{i.e.}\xspace}

\makeatletter
\newcommand{\linebreakand}{%
  \end{@IEEEauthorhalign}
  \hfill\mbox{}\par
  \mbox{}\hfill\begin{@IEEEauthorhalign}
}
\makeatother

\def\BibTeX{{\rm B\kern-.05em{\sc i\kern-.025em b}\kern-.08em
    T\kern-.1667em\lower.7ex\hbox{E}\kern-.125emX}}

\begin{document}

\title{%
Key Exchange in the Quantum Era: Evaluating a Hybrid System of Public-Key Cryptography and Physical-Layer Security
\thanks{This work was funded in part by the US National Science Foundation under grant ECCS-2029323 and by the Deutsche Forschungsgemeinschaft (DFG, German Research Foundation) under Germany’s Excellence Strategy - EXC 2092 CASA - 390781972.}
}

\author{\IEEEauthorblockN{  Paul Staat\IEEEauthorrefmark{1}, 
                            Meik Dörpinghaus\IEEEauthorrefmark{2}, 
                            Azadeh Sheikholeslami\IEEEauthorrefmark{3}, 
                            Christof Paar\IEEEauthorrefmark{1}, 
                            Gerhard Fettweis\IEEEauthorrefmark{2} and 
                            Dennis Goeckel\IEEEauthorrefmark{4}}
        \IEEEauthorblockA{\IEEEauthorrefmark{1}Max Planck Institute for Security and Privacy, Bochum, Germany, \{paul.staat, christof.paar\}@mpi-sp.org\\
                          \IEEEauthorrefmark{2}Vodafone Chair Mobile Communications Systems, Technische Universität Dresden, Dresden, Germany,\\ \{meik.doerpinghaus, gerhard.fettweis\}@tu-dresden.de\\
                          \IEEEauthorrefmark{3} Math and Computer Science Department, Suffolk University, Boston, USA, asheikholeslami@suffolk.edu\\
                          \IEEEauthorrefmark{4} Dept. Electrical and Computer Engineering, University of Massachusetts Amherst, USA, goeckel@ecs.umass.edu
}
}
    
\maketitle

\begin{acronym}

\acro{AES}{Advanced Encryption Standard}

\acro{LoS}{line of sight}

\acro{OFDM}{orthogonal frequency division multiplexing}

\acro{PCB}{printed circuit board}

\acro{RF}{radio frequency}

\acro{RSSI}{received signal strength indicator}

\acro{SNR}{signal-to-noise ratio}

\acro{JSR}{jamming-to-signal ratio}

\acro{SJNR}{signal-to-jamming-and-noise ratio}

\acro{TDD}{time-division duplex}

\acro{MIMO}{multiple-input and multiple-output}

\acro{ADC}{analog-to-digital converter}

\acro{RSA}{Rivest–Shamir–Adleman}

\acro{DH}{Diffie-Hellman}

\acro{NIST}{National Institute of Standards and Technology}

\acro{JKE}{jamming key exchange}

\acro{PHY}{physical layer}

\acro{PQC}{post-quantum cryptography}

\acro{SFDR}{spurious-free dynamic range}

\acro{TLS}{transport layer security}

\acro{DSP}{digital signal processing}

\acro{FODL}{fiber-optic delay line}

\acro{ENOB}{effective number of bits}

\acro{DAC}{digital-to-analog~converter}

\end{acronym}

\begin{abstract}
\bstctlcite{IEEEexample:BSTcontrol}
Today's information society relies on cryptography to achieve security goals such as confidentiality, integrity, authentication, and non-repudiation for digital communications. Here, public-key cryptosystems play a pivotal role to share encryption keys and create digital signatures. However, quantum computers threaten the security of traditional public-key cryptosystems as they can tame computational problems underlying the schemes, \ie, discrete logarithm and integer factorization. The prospective arrival of capable-enough quantum computers already threatens today's secret communication in terms of their long-term secrecy when stored to be later decrypted. Therefore, researchers strive to develop and deploy alternative schemes.

In this work, evaluate a key exchange protocol based on combining public-key schemes with physical-layer security, anticipating the prospect of quantum attacks. If powerful quantum attackers cannot \textit{immediately} obtain private keys, legitimate parties have a window of short-term secrecy to perform a physical-layer \acf{JKE} to establish a long-term shared secret. Thereby, the protocol constraints the computation time available to the attacker to break the employed public-key cryptography. In this paper, we outline the protocol, discuss its security, and point out challenges to be resolved.

\end{abstract}

\section{Introduction}

The introduction of public-key cryptography in the 1970s \cite{diffieNewDirectionsCryptography1976, rivestMethodObtainingDigital1978} was a milestone in the age of telecommunications. Marking the beginning of a new era, previously unprecedented applications such as key exchange over insecure channels and digital signatures have revolutionized secure communications: The earliest and most prominent \ac{DH} and \ac{RSA} cryptosystems found ubiquitous deployment with the rise of the Internet in the 1990s. Decades later, the schemes today are still used in major network security protocols such as \ac{TLS} %
and have substantially contributed to building today's connected world.

The security of \ac{RSA} and \ac{DH} public-key cryptosystems is based on the hardness of computational problems, \ie, solving integer factorization and discrete logarithm. Thus, Peter Shor's discovery of a quantum algorithm~\cite{shorAlgorithmsQuantumComputation1994} to solve both problems in polynomial-time caused a stir in the cryptographic community: \ac{RSA} and \ac{DH} become insecure once capable-enough quantum computers are available. In turn, researchers proclaimed a race against time to replace the schemes before quantum computing reaches maturity~\cite{bernsteinPostquantumCryptography2017}. Although \textit{quantum supremacy} has already been declared \cite{aruteQuantumSupremacyUsing2019}, current quantum computers are still far from being capable of breaking today's cryptography, demanding thousands of logical qubits and likely millions of physical qubits~\cite{gidneyHowFactor20482021, moscaCybersecurityEraQuantum2018}. The immediate threat may appear far off at first glance; however, the long-term secrecy of current communications is threatened by store-now-decrypt-later strategies. Consequently, we must consider potential threats beyond the capabilities of present-day technology, particularly given the uncertainties regarding future advances in quantum computing~\cite{moscaQuantumThreatTimeline2023}. Thus, the American \ac{NIST} recently selected a set of \textit{post-quantum} cryptographic algorithms for standardization~\cite{moodyTransitionPostQuantumCryptography2024}. 

Given the potential rise of quantum computing (along with continuous algorithmic progress~\cite{boudotStateArtInteger2022}), classical public-key algorithms may eventually be broken, making them ineffective in providing long-term secrecy. In this article, however, we provide a more optimistic view: %
We argue that classical public-key algorithms yield at least a \textit{temporal} advantage over a powerful adversary, providing an opportunity to achieve everlasting security. For example, Sheikholeslami~\etal~\cite{sheikholeslamiEverlastingSecrecyExploiting2013, sheikholeslamiJammingBasedEphemeral2015} have shown that legitimate parties can leverage a shared cryptographic secret to obtain everlasting security on the physical layer. Here, mutually known pseudo-random power modulation~\cite{sheikholeslamiEverlastingSecrecyExploiting2013} and jamming signals~\cite{sheikholeslamiJammingBasedEphemeral2015} allow to exploit imperfections of the eavesdropping receiver and diminish reliance on channel conditions. This approach was later extended to a scenario where a friendly radar acts as a trusted third party to generate jamming signals~\cite{guanAchievableInformationTheoreticSecrecy2019}.

Our idea is to use classical public-key cryptography, \eg, \ac{RSA} or \ac{DH}, to establish an initial secret key. In the subsequent \textit{\acf{JKE}} phase, building on previous work by Sheikholeslami~\etal~\cite{sheikholeslamiJammingBasedEphemeral2015}, the initial secret is used to mask (jam) the physical-layer exchange of a long-term (everlasting) secret by adding a pseudo-random jamming signal. A core assumption is that an eavesdropper cannot store the jammed receive signal for later processing without loss of information. Instead, the jamming signal needs to be removed in real-time which can only be accomplished when possessing the initial secret. Thus, the attacker 
has to find the initial secret before the \ac{JKE} finishes and is forced into a timing constraint, \ie, imposing requirements on the speed of future (quantum) attackers.

\subsection{Our Contribution}

We evaluate the idea of Sheikholeslami~\etal~\cite{sheikholeslamiJammingBasedEphemeral2015} to achieve everlasting security on the wireless radio physical layer when incorporated into a contemporary cryptographic system. Such a system holds promise to put stringent timing requirements on quantum attackers attempting to break classical public-key cryptography-based key exchange. The challenge is that such a system involves current and future assumptions on radio receiver operation, cryptographic protocols, quantum computing, and capabilities for the storage of analog signals, hence spanning a wide multidisciplinary range. In this paper, we define key metrics that will determine the utility of the system and provide estimates for key parameters grounded in current technology and future trajectories. We propose the notions of temporal advantages and non-storage channels to inform the design of cryptographic protocols. Additionally, we highlight key challenges that must be addressed, and identify further questions that are essential to understanding the system's utility and security promises.

\section{The Protocol}
In this work, we consider two legitimate parties Alice and Bob who seek to establish a shared symmetric long-term key $k_L$ with a length of at least 256~bit. Alice and Bob operate on the physical layer and are capable of processing analog signals, controlling waveforms transmitted over the wireless communication channel but also to encode and decode digital information. We assume a wiretap channel~\cite{wynerWiretapChannel1975} scenario where the passive eavesdropper Eve taps the communication channel between Alice and Bob. Eve's front end is constrained, limiting their ability to perfectly store the analog eavesdropped signal. However, Eve is capable of executing both classical and quantum attacks against public-key cryptographic primitives.

\begin{figure}
\centering
\includegraphics[width=1.0\linewidth]{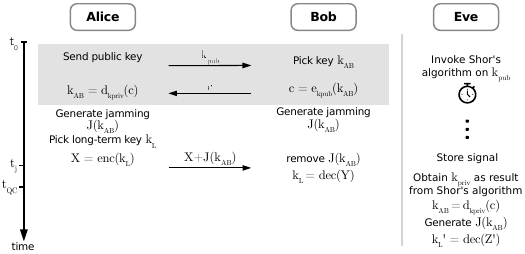}
\caption{Alice and Bob use classical public-key cryptography, \eg, \ac{RSA} or \ac{DH}, to establish a key $k_{AB}$ which enables their \ac{JKE} to establish $k_L$. Concurrently, Eve seeks to compute $k_{AB}$ and obtain $k_L$. Everlasting security of $k_L$ is given if $t_{j} < t_{QC}$.}
\label{fig:example_protocol}
\end{figure}

We propose a two-phase protocol: First, Alice and Bob use classical public-key cryptography to establish an initial shared secret. In the second phase, Alice and Bob use the initial secret as a seed to generate a jamming noise signal overshadowing the transmission of the analog signal that encodes the final long-term secret. \autoref{fig:example_protocol} shows the message flow of Alice and Bob, including the initial key transport and the \ac{JKE}. In the following, we elaborate both phases.

\subsection{Phase 1: Initial Key Establishment}

Alice and Bob begin by establishing a shared symmetric key $k_{AB}$. For this, they employ public-key cryptography in the same fashion as typical security protocols, \eg, \ac{TLS} when using the Internet. In~\autoref{fig:example_protocol}, the corresponding messages are highlighted by the area with gray background. Here, we consider an exemplary variant utilizing \ac{RSA} for key encapsulation\footnote{Alternatively, other public-key primitives such as the generalized Diffie-Hellman problem or more recent \ac{PQC} schemes~\cite{moodyTransitionPostQuantumCryptography2024} can be used.}: Alice has a pair of public and private keys $(k_{pub}, k_{priv})$ and first shares the public key with Bob. Then, Bob generates the symmetric 256-bit long key $k_{AB}$, encrypts it with Alice' public key, transmits the ciphertext to Alice who obtains $k_{AB}$ by decrypting with their private key. This is known as a key encapsulation mechanism.

The \ac{RSA} keys are denoted as $k_{pub} = (n,e)$ and $k_{priv} = d$. The public and private exponents $e$ and $d$ are used for encryption and decryption through modular exponentiation in the residual class $n$. Importantly, $n = p \cdot q$ is the product of two large prime numbers $p$ and $q$ and the security of \ac{RSA} is rooted in the inability to recover the two prime factors from $n$, \ie, integer factorization of large numbers.

\subsection{Phase 2: Jamming Key Exchange}

\begin{figure}
\centering
\includegraphics[width=0.85\linewidth]{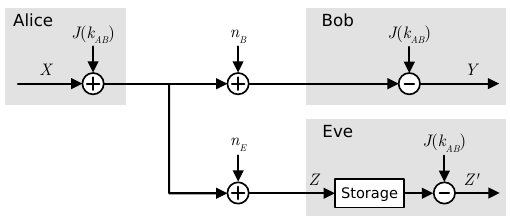}
\caption{\ac{JKE} in a wiretap setting. }
\label{fig:jamming_wiretap}
\end{figure}

In the second step, Alice and Bob perform a \ac{JKE}~\cite{sheikholeslamiJammingBasedEphemeral2015} to establish the long-term secret $k_L$. Alice picks $k_L$ and encodes it into a message $X$, comprised of symbols $X_n$, to be transmitted to Bob. Crucially, before transmission of $X$ over the public channel, Alice adds $J(k_{AB})$ which is a pseudo-random jamming signal. Each jamming symbol comprises of $w$ bits generated with a cryptographically secure pseudo-random number generator seeded from the initial secret $k_{AB}$. On the receiving side, Bob reproduces $J(k_{AB})$ and removes it from the received signal to eventually recover $k_L$ from the received signal~$Y$. The \ac{JKE} ends at time $t_{j}$. As shown in previous work by Sheikholeslami~\etal~\cite{sheikholeslamiJammingBasedEphemeral2015}, Eve loses information on $X$ when attempting to persistently store the superposition of $X$ and $J(k_{AB})$. We refer to this mechanism as a \textit{non-storage channel}.%

Following a (modified) Wyner wiretap channel setting~\cite{wynerWiretapChannel1975}, \autoref{fig:jamming_wiretap} depicts a block diagram comprising the previously outlined processing steps for the protocol. Here, we assume that Eve only possesses $k_{AB}$ after the \ac{JKE} has finished, enforcing storage of the signal.

\section{Security Analysis}
\label{sec:discussion}
In this section, we discuss the security of the protocol, elaborating on attack strategies involving breaking the employed public-key cryptography, storing the \ac{JKE} for later processing, and insufficient channel conditions.

\subsection{Temporal Aspects: Who Is Faster?}

For the attacker to be able to perform the very same processing like Bob -- avoiding persistent signal storage -- knowledge of $k_{AB}$ is required to reproduce and remove $J(k_{AB})$ on-the-fly upon signal reception. For this, Eve could attempt to break the employed public-key cryptography, calculating $k_{priv}$ from $k_{pub}$ to eventually obtain $k_{AB}$. Thus, the protocol constructs a race condition between Eve attempting to find $k_{AB}$ and the \ac{JKE}, imposing a lower bound on the attacker's speed. 

In the particular case of \ac{RSA}, Eve could make use of any factoring strategy, including quantum algorithms, and would fully recover $k_{AB}$ after time $t_{QC}$. Please note that we can neglect time for decapsulation of $k_{AB}$ from $c$ 
and generating $J(k_{AB})$ as this has to likewise be performed by the legitimate parties. %
If $t_{QC} > t_j$, \ie, the attacker would only gain knowledge of $k_{AB}$ after the \ac{JKE}, Alice and Bob enjoy a temporal advantage over Eve during the \ac{JKE}. In consequence, Eve would be unable to predict $J(k_{AB})$. Eve might succeed to partially recover $k_{AB}$ in time. However, given that $J$ is generated using a cryptographifcally secure pseudo-random generator, even single-bit deviations from the correct seed $k_{AB}$ would yield a completely different jamming signal, hence not aiding Eve in removing the actual jamming signal. %

As we estimate $t_j$ to lie deeply in the sub-second region (see~\autoref{sec:secrecy_rates}), Eve faces a significant hurdle to push $t_{QC}$ below $t_j$. Arguably, we cannot make projections on the future trajectory of $t_{QC}$, depending on unforeseeable technological progress in (quantum) computing and factoring algorithms. To date, the largest number factored using classical computing is 829~bit long, an effort that took 2,700~CPU~core~years~\cite{boudotStateArtInteger2022}. Yet, it was estimated that a hypothetical quantum computer with millions of noisy qubits could break \ac{RSA}-2048 within 8h~\cite{gidneyHowFactor20482021}. We refer the reader to \autoref{sec:quantum_discussion} for additional discussion on quantum computing.

\subsection{Jamming Key Exchange Signal Storage}

When Eve does not obtain $k_{AB}$ in time, \ie, $t_{QC} > t_{j}$, they must store the \ac{JKE} signal and follow a store-now-decrypt-later strategy. At this point, a key assumption comes into play: storing the \ac{JKE} signal implies a loss of information on $X$, laying the foundation for Alice and Bob to gain secrecy which they utilize to establish their long-term key $k_L$. Thus, the protocol's security depends on Eve's inability for lossless storage of the \ac{JKE} signal. In the following, we  discuss options for both analog and digital storage media.

\subsubsection{Digital storage}
\label{section:ad_technologies}

We assume Eve samples the eavesdropped \ac{JKE} signal $Z = X +J(k_{AB}) + n_E$ using an \ac{ADC}. Thus, Eve digitizes the message $X$ and the jamming signal $J(k_{AB})$, both of which contribute to the inherent quantization noise arising with the sampling process. Even though Eve may obtain $k_{AB}$ at a later point in time, allowing to remove $J(k_{AB})$ and obtain $Z'$, the quantization noise hinders Eve to perfectly recover $X$, \ie, a loss of information. 

The amount of noise due to sampling is determined by the employed \ac{ADC} hardware. In particular, as discussed in detail in Section~V of \cite{sheikholeslamiJammingBasedEphemeral2015}, \acp{ADC} are limited by aperture jitter which is the uncertainty of the sample time (caused by the \ac{ADC} hardware). It limits the product of the \ac{ADC} amplitude resolution (\ie, \ac{ENOB}) and bandwidth of the signal to be converted. In this regard, \cite{sheikholeslamiJammingBasedEphemeral2015} has shown that Eve can maximize the mutual information between $Z'$ and $X$ by minimizing the aperture jitter of their \ac{ADC}. However, with today's state-of-the-art \acp{ADC}, offering about 50~fs~rms aperture jitter, the jamming signal can be designed such that Eve cannot store the \ac{JKE} signal with sufficient fidelity (see~\autoref{sec:secrecy_rates}).

\subsubsection{Analog storage}

Alternatively, Eve could employ a form of analog storage, being equivalent to delay (\enquote{hold}) the signal without loss of fidelity while attempting to break $k_{AB}$ and then remove the jamming signal. However, delaying wideband signals is considered challenging to realize from a technical perspective. %
Emerging from the electronic warfare context, \acp{FODL} recirculate \ac{RF} signals on a loop of optical fiber to realize delay~\cite{diehlMicrowavePhotonicDelay2015}. However, \acp{FODL} offer rather short delays in the order of micro- to milliseconds while still degrading fidelity of the signal. Thus, Eve would only gain marginal computing time while potentially sacrificing their mutual information. Apart from \acp{FODL}, traditional analog storage media such as tubes, tapes, or disks could possibly be utilized. However, we are not aware of any references attempting using such technology to store analog wideband \ac{RF} signals or reporting noise figure and \ac{SFDR} performance. Thus, we conclude that digital storage currently is the most promising approach to store the \ac{JKE} signal, motivating the analysis presented in the next section.

\subsection{Conditions for Positive Secrecy Rates} 
\label{sec:secrecy_rates}

To achieve positive secrecy rates, classical physical-layer security for the wiretap channel demands the legitimate channel quality to be better than that of the eavesdropper. This assumption grants the malicious adversary significant room to optimize their channel, \eg, strategically positioning very close to Alice. Importantly, the \ac{JKE}-based approach provides an opportunity to decouple secrecy rates from channel conditions.

\begin{figure}[!t]
\centering
    \begin{subfigure}{0.8\columnwidth}
        \includegraphics[width=1\linewidth,trim={0.1cm 1.2cm 0.6cm 2.0cm},clip]{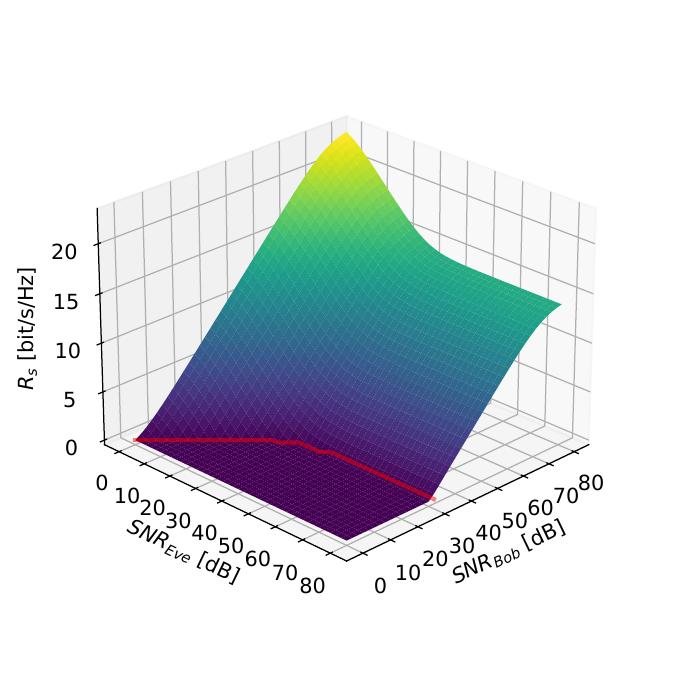}
        \caption{}
        \label{fig:secrecy_rates_vs_SNR}
    \end{subfigure}%
    \\
    \begin{subfigure}{0.8\columnwidth}
        \includegraphics[width=1\linewidth,trim={0.1cm 0.9cm 0.6cm 1.5cm},clip]{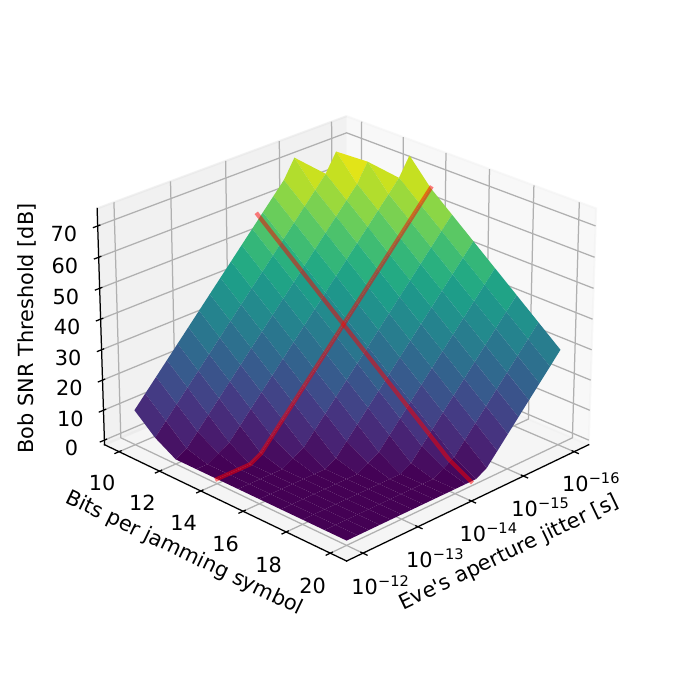}
        \caption{}
        \label{fig:positive_rs_threshold}
    \end{subfigure}%
    \caption{\ac{JKE} secrecy rate evaluation for a 40~MHz signal bandwidth and 500~fs~rms aperture jitter of Bob's \ac{ADC}. (a)~Secrecy rate versus \ac{SNR} of Eve's channel and \ac{SNR} of Bob's channel for 5~fs aperture jitter of Eve's \ac{ADC} and 14 bits per jamming symbol. 
    (b)~Minimum \ac{SNR} of Bob to achieve a positive secrecy rate versus the number of jamming bits per symbol and the aperture jitter of Eve's \ac{ADC} when Eve's channel is noiseless.}
    \label{fig:secrecy_rate_surfaces}%
\end{figure}
\par %

For the considered wiretap channel with additive jamming inserted by Alice which is known to Bob but unknown to Eve, the secrecy rate has been evaluated in \cite{sheikholeslamiJammingBasedEphemeral2015}. As in \cite{sheikholeslamiJammingBasedEphemeral2015} we assume that Bob and Eve are using \acp{ADC} with uniform quantization having $b_{\textrm{B}}$ and $b_{\textrm{E}}$ bits amplitude resolution, respectively. Using the optimal dynamic range, this results in a quantization resolution for Bob of  
\begin{equation}
\delta_{\textrm{B}}=\frac{2l\sqrt{P}}{2^{b_{\textrm{B}}}},
\end{equation}
where $P=\mathrm{E}[X_n^2]$, \ie, the power of the symbols $X_n$ from the message $X$, constituting the useful information-carrying component of the transmit signal, and $l=2.5$ is a numerically optimized parameter to maximize the mutual information, see~\cite{sheikholeslamiEverlastingSecrecyExploiting2013, sheikholeslamiJammingBasedEphemeral2015}.

As Eve is not able to remove the additive jamming signal before A/D conversion, they will optimize the dynamic range of the ADC to minimize the resulting quantization distortion for the sum of the useful transmit signal $X$ and the additive jamming signal $J(k_{\textrm{AB}})$. This results in a quantization resolution
\begin{equation}
\delta_{\textrm{E}}=\frac{2l\sqrt{P}}{2^{b_{\textrm{E}}-w}},
\end{equation}
with $w$ being the number of bits per jamming symbol. For this setting it has been shown in \cite{sheikholeslamiJammingBasedEphemeral2015} that the secrecy rate is lower-bounded by
\begin{equation}
    R_s = W\, \left[ \textrm{log}\left( \frac{P + \sigma^2_{\textrm{B}} + \frac{\delta^2_{\textrm{B}}}{12}}{\sigma^2_{\textrm{B}} + \frac{\delta^2_{\textrm{B}}}{12}} \right) -  \textrm{log}\left( \frac{P + \sigma^2_{\textrm{E}} + \frac{\delta^2_{\textrm{E}}}{12}}{\sigma^2_{\textrm{E}} + \frac{\delta^2_{\textrm{E}}}{2 \pi e}} \right) \right],\label{eq::eq12}
\end{equation}
where $\sigma^2_{\textrm{B}}$ and $\sigma^2_{\textrm{E}}$ are the variances of the additive white Gaussian noise processes $n_{\textrm{B}}$ and $n_{\textrm{E}}$ of the channels to Bob and to Eve, respectively. $W$ is the bandwidth of the channel.

For conservative security estimates, we assume that Eve has access to advanced \ac{ADC} technology, offering a bandwidth-amplitude resolution product one order of magnitude beyond today's state-of-the-art, manifesting itself as 5~fs~rms aperture jitter. Anticipating significant capability imbalance, we assume Bob's \ac{ADC} to be two orders of magnitude behind, having 500~fs~rms aperture jitter. From this, we can determine the \ac{ADC} amplitude resolutions $b_{\textrm{B}}$ and $b_{\textrm{E}}$ as the respective \acp{ENOB}. For a given aperture jitter $t_{aj}$ and signal bandwidth $W$, the achievable amplitude resolution in terms of the \ac{ENOB} is given by \cite{sheikholeslamiJammingBasedEphemeral2015}
\begin{equation}
\textrm{ENOB}=-\frac{20\log_{10}(2\pi W t_{aj})+1.76}{6.02}.
\end{equation}

\autoref{fig:secrecy_rates_vs_SNR} shows the secrecy rate achieved by Alice and Bob as a function of the respective \acp{SNR} of Bob and Eve when Alice uses 14~bits per jamming symbol. Based on the secrecy rate result, we can estimate the \ac{JKE} duration $t_j$. For example, when Bob and Eve have respective \acp{SNR} of 32~dB and 80~dB, assuming a conservative protocol efficiency of 0.1\%, the transmission of $k_L$, comprising of 256 secret bits, would take approx. 11.52~ms. The red line in the plot indicates the region where $R_s$ becomes positive. Importantly, we can see that the secrecy rate rises quickly once a given minimum \ac{SNR} at Bob is reached, even when Eve enjoys very high \ac{SNR}. Hence, for certain constellations, the \ac{JKE} enables secret communication regardless of Eve's location and channel quality. Thus, \autoref{fig:positive_rs_threshold} shows the minimum required \ac{SNR} of Bob to achieve positive secrecy when Eve has a significant \ac{SNR} advantage over Bob, \ie, directly tapping Alice' transmissions such that $\textrm{SNR}_{Eve}= \infty$. We plot this \ac{SNR} threshold over the number of bits per jamming symbol $w$ and the \ac{ADC} aperture jitter of Eve. Hence, this plot allows to assess the protocol's channel-independent security promises in light of technology parameters, namely \ac{ADC} quality and analog interference cancellation capabilities at Bob (see~\autoref{sec::interference_cancellation}). The red lines indicate the technology levels from the previous plot, \ie, 14 bits per jamming symbol and 5~fs~rms aperture jitter at Eve. Moreover, we can see that the number of jamming bits per symbol is a security parameter that allows to inflict increased aperture jitter requirements on Eve.

\subsection{Everlasting Security}
The original proposal by Sheikholeslami~\etal~\cite{sheikholeslamiJammingBasedEphemeral2015} was intended for confidential communication with everlasting security. In a quantum era, this would still work with the \ac{JKE}, yet demanding periodic re-keying to reset the attack time race condition. Please note that the \ac{JKE} proposal is meant to establish a shared symmetric key $k_L$ to be used for cryptographic symmetric encryption. This key, however, only has everlasting security when it remains unused. That is, a computationally unbounded adversary would be capable of breaking symmetric encryption, effectively determining $k_L$. Still, reflecting on the quantum threat, it is expected that symmetric encryption schemes like \ac{AES} will remain secure since Grover's algorithm provides only a quadratic speedup~\cite{moscaQuantumThreatTimeline2023}. That is, the security level of a 256-bit encryption would be reduced to $2^{128}$.

\section{Technology Discussion}

\subsection{ADC Technology Trends}
As shown in the previous section, future technology improvement in terms of \ac{ADC} aperture jitter performance would allow Eve to digitally store \ac{JKE} signals and diminish positive secrecy rate conditions. However, several works~\cite{sheikholeslamiJammingBasedEphemeral2015, jonssonSurveyDConverterPerformance2010, murmannRaceExtraDecibel2015} anticipate an eventual saturation of \ac{ADC} aperture jitter performance. Murmann~\cite{murmannRaceExtraDecibel2015} pointed out that progress is likely continuing to be incremental, yet at some point will rather be limited by the impurity of clocks. Data from \cite{murmannADCPerformanceSurvey}, collecting \ac{ADC} designs presented at the area conferences VLSI and ISSCC, suggests that bandwidth-amplitude resolution performance doubled approx. every 4.57 years between 2005 and 2024 (using the same analysis as in \cite{murmannRaceExtraDecibel2015}). Conservatively assuming this trend to continue further with current state-of-the-art being at 50~fs rms jitter, the 5~fs rms jitter we consider in this paper would be reached around~2040. However, between 2010 and 2024, performance doubled only approx. every 8.34 years. Finally, we would like to point out that photonic solutions might further push the aperture jitter boundaries~\cite{khiloPhotonicADCOvercoming2012} and should be taken into account for future assessments.

\subsection{Interference Cancellation}
\label{sec::interference_cancellation}

While the number of bits per jamming symbol is a security parameter, it likewise is a technology parameter. That is, the jamming signal resolution dictates the minimum performance of Bob's interference cancellation. In particular, Bob has to remove the very strong (but known) jamming signal in the analog domain, ideally without leaving a residual. In \cite{guanAchievableInformationTheoreticSecrecy2019}, when a similar approach was employed with a co-channel radar taking the role of the jammer, residual interference was modeled, showing that this degrades the available secrecy rate.

Fortunately, the cancellation of known interfering signals has seen significant study as it is an enabling technology of full-duplex radios, allowing simultaneous transmission and reception in the same band. In such a system, the locally-generated transmit signal can be, for instance, 120~dB above the signal of interest received from a distant transmitter. Therefore, analog cancellation techniques are employed to reduce the co-channel interference, \eg, by 63~dB~\cite{bharadiaFullDuplexRadios2013} and 54~dB~\cite{nagulu66FullDuplexReceiver2021} across an 80~MHz bandwidth. As discussed in~\cite{bharadiaFullDuplexRadios2013}, cancellation performance can be further enhanced by approx.~30~dB~when additionally employing digital techniques. With the standard conversion of 6~dB per bit, this means we can expect Bob to be able to remove signals with roughly 14 to 16 bits per symbol resolution. However, owing to the remote nature of the legitimate parties, the receiving side must tightly synchronize to fully realize such cancellation performance. This challenge is acknowledged by recent experimental studies of known remote interference cancellation~\cite{parlinKnownInterferenceCancellationCooperative2023, dotyAnalogCancellationKnown2024}.

\subsection{Transmit Signal Synthesis}

One potential concern about the system is that the transmitter Alice must build a signal that has high dynamic range: Alice's small signal $X$ embedded in the large jamming signal $J(k_{AB})$, suggesting at first glance the requirement of a $b_B+w$ bit \ac{DAC}. Although \acp{DAC} are much easier to build than \acp{ADC}, this could still prove to be a daunting task for some parameter settings.  However, such a \ac{DAC} is not required.  Because Alice's voltage levels need not be aligned with the jammer's voltage levels, Alice's and the jammer's signal can be converted to analog separately and then summed.  Hence, only two \acp{DAC} of moderate resolution are required at Alice.  Similarly, Bob only needs to generate the jamming signal to effect cancellation, hence requiring a $w$-bit \ac{DAC}.

\subsection{Evolution of Quantum Computing}
\label{sec:quantum_discussion}
Current quantum computers do not immediately threaten cryptographic primitives (yet future quantum computers could affect secrecy of today's data). While Shor's algorithm has been experimentally demonstrated~\cite{vandersypenExperimentalRealizationShors2001, amicoExperimentalStudyShors2019, martin-lopezExperimentalRealizationShors2012}, this was limited to factoring of rather small numbers such as $n = 15, 21, 35$, in some instances even assuming knowledge on the solution~\cite{smolinOversimplifyingQuantumFactoring2013}. However, it was estimated that a hypothetical quantum computer with millions of noisy qubits could break \ac{RSA}-2048 within 8h~\cite{gidneyHowFactor20482021}. Making statements whether quantum computers will ever scale sufficiently to become cryptographically relevant is clearly difficult. Still, an annual survey among leading experts in the field~\cite{moscaQuantumThreatTimeline2023} shows that this particular cohort tends towards expecting such scale -- considering the threat of breaking RSA-2048 within 24~hours -- could be reached within 30 years. %

\section{Conclusion}
In the present paper we evaluated a hybrid key exchange protocol, combining public-key cryptography and physical-layer security. In particular, we investigated the idea of using public-key cryptography to obtain a temporal advantage over a passive eavesdropper. Using additive jamming signals, this advantage allows constructing a physical-layer non-storage channel to eventually establish a long-term secret. The secrecy rate of this channel is directly related to the eavesdropper's information loss during analog-to-digital conversion of the physical-layer signals for later processing, rooted in finite quantization resolution of the \ac{ADC}.

Using a (short) computational advantage to enable exploitation of physical phenomena which inflict information loss on the eavesdropper is an interesting idea, enabling communication with everlasting security. While such a system could allow formulation of speed requirements to attacks against public-key cryptographic primitives, we caution the reader to emphasize the system's inherent interconnection with technology. That is, future concurrent progress on quantum computing, storage of analog signals, and interference cancellation must be closely examined to re-assess whether the protocol can uphold its security promises. Therefore, we believe that the next logical step would be a more formal security analysis of the protocol and designing additional physical non-storage mechanisms enabled through computational advantages.

\subsection{Future Work}
The limitation of the bandwidth-amplitude resolution product of \acp{ADC} can also be exploited by using the jamming signal to increase the bandwidth of the receive signal of the eavesdropper. In this regard,  \cite{sheikholeslamiJammingBasedEphemeral2015} considered frequency hopping for bandwidth expansion. While this approach enables significantly increased secrecy rates, it has the drawback that at each time instant the received signal still has the original bandwidth, the eavesdropper just does not know which frequency band is momentarily used. It can be imagined that the eavesdropper uses a bank of narrowband \acp{ADC} running in parallel, thus, enabling digitization of the individual subbands without gaining additional secrecy in comparison to the additive jamming approach. This could be avoided by using a multiplicative jamming signal allowing to expand the signal bandwidth so far that the eavesdropper will be unable to digitize this wideband signal without a significant decrease in the amplitude resolution which is governed by the limited bandwidth-amplitude product. We are currently studying the implications of such multiplicative jamming processes.

\bibliographystyle{IEEEtran2}
\bibliography{temp_control, tempadv_refs_final}

\begin{thebibliography}{10}
\providecommand{\url}[1]{#1}
\csname url@samestyle\endcsname
\providecommand{\newblock}{\relax}
\providecommand{\bibinfo}[2]{#2}
\providecommand{\BIBentrySTDinterwordspacing}{\spaceskip=0pt\relax}
\providecommand{\BIBentryALTinterwordstretchfactor}{4}
\providecommand{\BIBentryALTinterwordspacing}{\spaceskip=\fontdimen2\font plus
\BIBentryALTinterwordstretchfactor\fontdimen3\font minus
  \fontdimen4\font\relax}
\providecommand{\BIBforeignlanguage}[2]{{%
\expandafter\ifx\csname l@#1\endcsname\relax
\typeout{** WARNING: IEEEtran.bst: No hyphenation pattern has been}%
\typeout{** loaded for the language `#1'. Using the pattern for}%
\typeout{** the default language instead.}%
\else
\language=\csname l@#1\endcsname
\fi
#2}}
\providecommand{\BIBdecl}{\relax}
\BIBdecl

\bibitem{diffieNewDirectionsCryptography1976}
\BIBentryALTinterwordspacing
W.~Diffie and M.~Hellman, ``New directions in cryptography,'' \emph{IEEE Trans.
  Inf. Theory}, vol.~22, no.~6, pp. 644--654, Nov. 1976.
\BIBentrySTDinterwordspacing

\bibitem{rivestMethodObtainingDigital1978}
\BIBentryALTinterwordspacing
R.~L. Rivest, A.~Shamir, and L.~Adleman, ``A method for obtaining digital
  signatures and public-key cryptosystems,'' \emph{Commun. ACM}, vol.~21,
  no.~2, pp. 120--126, Feb. 1978.
\BIBentrySTDinterwordspacing

\bibitem{shorAlgorithmsQuantumComputation1994}
\BIBentryALTinterwordspacing
P.~Shor, ``Algorithms for quantum computation: discrete logarithms and
  factoring,'' in \emph{Proc. 35th {{Annual Symposium}} on {{Foundations}} of
  {{Computer Science}}}, Nov. 1994, pp. 124--134.
\BIBentrySTDinterwordspacing

\bibitem{bernsteinPostquantumCryptography2017}
\BIBentryALTinterwordspacing
D.~J. Bernstein and T.~Lange, ``Post-quantum cryptography,'' \emph{Nature},
  vol. 549, no. 7671, pp. 188--194, Sep. 2017.
\BIBentrySTDinterwordspacing

\bibitem{aruteQuantumSupremacyUsing2019}
\BIBentryALTinterwordspacing
F.~Arute, K.~Arya \emph{et~al.}, ``Quantum supremacy using a programmable
  superconducting processor,'' \emph{Nature}, vol. 574, no. 7779, pp. 505--510,
  Oct. 2019.
\BIBentrySTDinterwordspacing

\bibitem{gidneyHowFactor20482021}
\BIBentryALTinterwordspacing
C.~Gidney and M.~Eker{\aa}, ``How to factor 2048 bit {{RSA}} integers in 8
  hours using 20 million noisy qubits,'' \emph{Quantum}, vol.~5, p. 433, Apr.
  2021.
\BIBentrySTDinterwordspacing

\bibitem{moscaCybersecurityEraQuantum2018}
\BIBentryALTinterwordspacing
M.~Mosca, ``Cybersecurity in an era with quantum computers: Will we be ready?''
  \emph{IEEE Secur. Priv.}, vol.~16, no.~5, pp. 38--41, Sep. 2018.
\BIBentrySTDinterwordspacing

\bibitem{moscaQuantumThreatTimeline2023}
\BIBentryALTinterwordspacing
M.~Mosca and M.~Piani, ``Quantum threat timeline report 2023,'' Tech. Rep.,
  2023. [Online]. Available:
  \url{https://globalriskinstitute.org/publication/2023-quantum-threat-timeline-report/}
\BIBentrySTDinterwordspacing

\bibitem{moodyTransitionPostQuantumCryptography2024}
\BIBentryALTinterwordspacing
D.~Moody, R.~Perlner \emph{et~al.}, ``Transition to post-quantum cryptography
  standards,'' {National Institute of Standards and Technology}, Gaithersburg,
  MD, Tech. Rep. NIST IR 8547 ipd, 2024. [Online]. Available:
  \url{https://nvlpubs.nist.gov/nistpubs/ir/2024/NIST.IR.8547.ipd.pdf}
\BIBentrySTDinterwordspacing

\bibitem{boudotStateArtInteger2022}
\BIBentryALTinterwordspacing
F.~Boudot, P.~Gaudry \emph{et~al.}, ``The state of the art in integer factoring
  and breaking public-key cryptography,'' \emph{IEEE Secur. Priv.}, vol.~20,
  no.~2, pp. 80--86, Mar. 2022.
\BIBentrySTDinterwordspacing

\bibitem{sheikholeslamiEverlastingSecrecyExploiting2013}
\BIBentryALTinterwordspacing
A.~Sheikholeslami, D.~Goeckel, and H.~{Pishro-Nik}, ``Everlasting secrecy by
  exploiting non-idealities of the eavesdropper's receiver,'' \emph{IEEE J.
  Sel. Areas Commun.}, vol.~31, no.~9, pp. 1828--1839, Sep. 2013.
\BIBentrySTDinterwordspacing

\bibitem{sheikholeslamiJammingBasedEphemeral2015}
\BIBentryALTinterwordspacing
------, ``Jamming based on an ephemeral key to obtain everlasting security in
  wireless environments,'' \emph{IEEE Trans. Wirel. Commun.}, vol.~14, no.~11,
  pp. 6072--6081, Nov. 2015.
\BIBentrySTDinterwordspacing

\bibitem{guanAchievableInformationTheoreticSecrecy2019}
\BIBentryALTinterwordspacing
B.~Guan and D.~L. Goeckel, ``Achievable information-theoretic secrecy in the
  presence of a radar,'' in \emph{Proc. {{MILCOM}} 2019 - 2019 {{IEEE Military
  Communications Conference}} ({{MILCOM}})}, Norfolk, VA, USA, Nov. 2019, pp.
  476--481.
\BIBentrySTDinterwordspacing

\bibitem{wynerWiretapChannel1975}
\BIBentryALTinterwordspacing
A.~D. Wyner, ``The wire-tap channel,'' \emph{Bell Syst. Tech. J.}, vol.~54,
  no.~8, pp. 1355--1387, Oct. 1975.
\BIBentrySTDinterwordspacing

\bibitem{diehlMicrowavePhotonicDelay2015}
\BIBentryALTinterwordspacing
J.~F. Diehl, J.~M. Singley \emph{et~al.}, ``Microwave photonic delay line
  signal processing,'' \emph{Appl. Opt.}, vol.~54, no.~31, p. F35, Nov. 2015.
\BIBentrySTDinterwordspacing

\bibitem{jonssonSurveyDConverterPerformance2010}
\BIBentryALTinterwordspacing
B.~E. Jonsson, ``A survey of {{A}}/{{D}}-converter performance evolution,'' in
  \emph{Proc. 2010 17th {{IEEE International Conference}} on {{Electronics}},
  {{Circuits}} and {{Systems}}}, Dec. 2010, pp. 766--769.
\BIBentrySTDinterwordspacing

\bibitem{murmannRaceExtraDecibel2015}
\BIBentryALTinterwordspacing
B.~Murmann, ``The race for the extra decibel: A brief review of current adc
  performance trajectories,'' \emph{IEEE Solid-State Circuits Mag.}, vol.~7,
  no.~3, pp. 58--66, 2015.
\BIBentrySTDinterwordspacing

\bibitem{murmannADCPerformanceSurvey}
\BIBentryALTinterwordspacing
------, ``{{ADC}} performance survey 1997-2024.'' [Online]. Available:
  \url{https://github.com/bmurmann/ADC-survey}
\BIBentrySTDinterwordspacing

\bibitem{khiloPhotonicADCOvercoming2012}
\BIBentryALTinterwordspacing
A.~Khilo, S.~J. Spector \emph{et~al.}, ``Photonic {{ADC}}: overcoming the
  bottleneck of electronic jitter,'' \emph{Opt. Express}, vol.~20, no.~4, p.
  4454, Feb. 2012.
\BIBentrySTDinterwordspacing

\bibitem{bharadiaFullDuplexRadios2013}
\BIBentryALTinterwordspacing
D.~Bharadia, E.~McMilin, and S.~Katti, ``Full duplex radios,'' in \emph{Proc.
  of the {{ACM SIGCOMM}} 2013 {{Conference}} on {{SIGCOMM}}}, ser. {{SIGCOMM}}
  '13.\hskip 1em plus 0.5em minus 0.4em\relax NY, NY, USA: ACM, 2013, pp.
  375--386.
\BIBentrySTDinterwordspacing

\bibitem{nagulu66FullDuplexReceiver2021}
\BIBentryALTinterwordspacing
A.~Nagulu, S.~Garikapati \emph{et~al.}, ``6.6 full-duplex receiver with
  wideband multi-domain fir cancellation based on stacked-capacitor, {{N}}-path
  switched-capacitor delay lines achieving {$>$}{{54dB SIC}} across {{80MHz
  BW}} and {$>$}{{15dBm TX}} power-handling,'' in \emph{Proc. 2021 {{IEEE
  International Solid-State Circuits Conference}} ({{ISSCC}})}, San Francisco,
  CA, USA, Feb. 2021, pp. 100--102.
\BIBentrySTDinterwordspacing

\bibitem{parlinKnownInterferenceCancellationCooperative2023}
\BIBentryALTinterwordspacing
K.~P{\"a}rlin, T.~Riihonen \emph{et~al.}, ``Known-interference cancellation in
  cooperative jamming: Experimental evaluation and benchmark algorithm
  performance,'' \emph{IEEE Wirel. Commun. Lett.}, vol.~12, no.~9, pp.
  1598--1602, Sep. 2023.
\BIBentrySTDinterwordspacing

\bibitem{dotyAnalogCancellationKnown2024}
\BIBentryALTinterwordspacing
J.~M. Doty, R.~W. Jackson, and D.~L. Goeckel, ``Analog cancellation of a known
  remote interference: Hardware realization and analysis,'' \emph{IEEE Wirel.
  Commun. Lett.}, vol.~13, no.~3, pp. 829--833, Mar. 2024.
\BIBentrySTDinterwordspacing

\bibitem{vandersypenExperimentalRealizationShors2001}
\BIBentryALTinterwordspacing
L.~M.~K. Vandersypen, M.~Steffen \emph{et~al.}, ``Experimental realization of
  {{Shor}}'s quantum factoring algorithm using nuclear magnetic resonance,''
  \emph{Nature}, vol. 414, no. 6866, pp. 883--887, Dec. 2001.
\BIBentrySTDinterwordspacing

\bibitem{amicoExperimentalStudyShors2019}
\BIBentryALTinterwordspacing
M.~Amico, Z.~H. Saleem, and M.~Kumph, ``Experimental study of {{Shor}}'s
  factoring algorithm using the {{IBM Q}} experience,'' \emph{Phys. Rev. A},
  vol. 100, no.~1, p. 012305, Jul. 2019.
\BIBentrySTDinterwordspacing

\bibitem{martin-lopezExperimentalRealizationShors2012}
\BIBentryALTinterwordspacing
E.~{Mart{\'i}n-L{\'o}pez}, A.~Laing \emph{et~al.}, ``Experimental realization
  of {{Shor}}'s quantum factoring algorithm using qubit recycling,'' \emph{Nat.
  Photonics}, vol.~6, no.~11, pp. 773--776, Nov. 2012.
\BIBentrySTDinterwordspacing

\bibitem{smolinOversimplifyingQuantumFactoring2013}
\BIBentryALTinterwordspacing
J.~A. Smolin, G.~Smith, and A.~Vargo, ``Oversimplifying quantum factoring,''
  \emph{Nature}, vol. 499, no. 7457, pp. 163--165, Jul. 2013.
\BIBentrySTDinterwordspacing

\end{thebibliography}

\end{document}